# Построение фермионного вакуума и фермионных операторов рождения и уничтожения в теории алгебраических спиноров

В.В.Монахов

*Санкт-Петербургский государственный университет,
198504, Россия, Санкт-Петербург, ул.Ульяновская, 1, v.v.monahov@spbu.ru*

В комплексных модулях над вещественными алгебрами Клиффорда четной размерности введены фермионные переменные – аналог базиса Витта. На их основе построены примитивные идемпотенты, являющиеся эквивалентными клиффордовыми вакуумами. Показано, что модули алгебр раскладывается на прямые суммы минимальных левых идеалов, порождаемых этими идемпотентами, и что фермионные переменные могут рассматриваться как математические объекты, более фундаментальные, чем спиноры.

We introduced fermionic variables in complex modules over real Clifford algebras of even dimension which are analog of the Witt basis. We built primitive idempotents which are a set of equivalent Clifford vacuums. It is shown that the modules are decomposed into direct sum of minimal left ideals generated by these idempotents and that the fermionic variables can be considered as more fundamental mathematical objects than spinors.

Рассмотрим комплексный модуль над вещественной алгеброй Клиффорда четной размерности $n = 2m$ сигнатуры $(p,q)$, $n = p + q$. Разобьем базис $e_\alpha$ генераторов алгебры на $m$ пар попарно: $e_1$ с $e_2$, $e_3$ с $e_4$, …, $e_{2m-1}$ с $e_{2m}$. Введем величины $s_\alpha = 1$ при $(e_\alpha)^2 = 1$, $s_\alpha = i$ при $(e_\alpha)^2 = -1$. Тогда $(s_\alpha)^2 = (e_\alpha)^2$. При этом $\{e_\alpha, e_\beta\} = (s_\alpha)^2 \delta^\alpha_\beta$. Здесь и далее, за исключением формулы (4), по повторяющимся индексам нет суммирования.

Введем переменные $E_\alpha$ такие, что $e_\alpha = s_\alpha E_\alpha$, $E_\alpha = (s_\alpha)^{-1} e_\alpha$, $(E_\alpha)^2 = 1$ для всех значений $\alpha$, и переменные $\theta^\alpha$ и $\overline{\theta}_\alpha$, которые мы будем называть фермионными:

$$\theta^\alpha = \frac{E_{2\alpha-1} - iE_{2\alpha}}{2}, \quad \overline{\theta}_\alpha = \frac{E_{2\alpha-1} + iE_{2\alpha}}{2}. \qquad (1)$$

Из (1) следует, что $(\theta^\alpha)^2 = (\overline{\theta}_\alpha)^2 = 0$, $\{\theta^\alpha, \overline{\theta}_\beta\} = \delta^\alpha_\beta$. В комплексных алгебрах Клиффорда $Cl(2m)$ величины, аналогичные $\theta^\alpha$ и $\overline{\theta}_\alpha$, называются базисом Витта [1],[2]. Но базисом Витта преобразуется с помощью ортогональных вращений, а фермионные переменные – с помощью псевдоортогональных, сохраняющих $(e_\alpha)^2$ для всех $\alpha$.

Матричные представления $\widetilde{\theta}^\alpha$ и $\widetilde{\overline{\theta}}_\alpha$ фермионных переменных $\theta^\alpha$ и $\overline{\theta}_\alpha$ строятся на основе (1) из гамма-матриц. Для одной пары это $\begin{pmatrix} 0 & 0 \\ 1 & 0 \end{pmatrix}$ и $\begin{pmatrix} 0 & 1 \\ 0 & 0 \end{pmatrix}$. Для более высоких размерностей алгебр Клиффорда используем алгоритм [2] построения гамма-матриц в пространствах более высокой размерности на основе прямого произведения матриц $2 \times 2$. Такую форму матриц фермионных переменных назовем канонической.

Алгебраические спиноры – элементы левого идеала, получающегося в результате умножения элементов клиффордова пространства на примитивный идемпотент. В качестве такого идемпотента используем

$$I_V = \overline{\theta}_1 \theta^1 \overline{\theta}_2 \theta^2 ... \overline{\theta}_m \theta^m. \qquad (2)$$

Поскольку $(I_V)^2 = (\overline{\theta}_1 \theta^1)^2 (\overline{\theta}_2 \theta^2)^2 ... (\overline{\theta}_m \theta^m)^2 = I_V$, то $I_V$, действительно, идемпотент. Множители вида $\overline{\theta}_\alpha \theta^\alpha$ в (2) коммутируют и эрмитово сопряженные, поэтому $(\overline{\theta}_1 \theta^1 \overline{\theta}_2 \theta^2 ... \overline{\theta}_m \theta^m)^+ = (\overline{\theta}_m \theta^m)...(\overline{\theta}_2 \theta^2)(\overline{\theta}_1 \theta^1) = \overline{\theta}_1 \theta^1 \overline{\theta}_2 \theta^2 ... \overline{\theta}_m \theta^m$, т.е. он эрмитово сопряженный, $I_V^+ = I_V$. Докажем, что $I_V$ примитивен. Пусть это не так, и $I_V$ состоит из суммы ортогональных идемпотентов $I_V = I_1 + I_2$, где $(I_1)^2 = I_1$, $(I_2)^2 = I_2$, $I_1 I_2 = I_2 I_1 = 0$. Тогда $I_1 I_V = I_1(I_1 + I_2) = I_1$, $I_V I_1 = (I_1 + I_2) I_1 = I_1$. При этом $\overline{\theta}_k I_1 = \overline{\theta}_k I_V I_1 = 0$, то есть $\overline{\theta}_k I_1 = 0$ для каждого $k$. Пусть $I_1 = A_1 + \overline{\theta}_k A_2$, где $A_1$ и $A_2$ — элементы модуля, не содержащие мономов с элементом $\overline{\theta}_k$, где в мономах все элементы $\overline{\theta}_j$ расположены слева от $\theta^l$. Тогда из $\overline{\theta}_k I_1 = 0$ следует, что $\overline{\theta}_k A_1 = 0$, что возможно только при $A_1 = 0$. Следовательно, $I_1 = \overline{\theta}_k A_2$. Это верно для каждого $k$, поэтому $I_1 = \overline{\theta}_1 \overline{\theta}_2 ... \overline{\theta}_m B$, где $B$ — элемент алгебры, не содержащий мономов с элементами $\overline{\theta}_k$. Аналогично, поскольку $I_1 I_V = I_1$, то $I_1 \theta^k = I_1 I_V \theta^k = 0$, $I_1 \theta^k = 0$ для каждого $k$, и мы получаем, что $I_1 = c_1 \overline{\theta}_1 \overline{\theta}_2 ... \overline{\theta}_m \theta^1 \theta^2 ... \theta^m$, где $c_1$ — комплексная константа. То есть $I_1$ отличается от $I_V$ только числовым коэффициентом $c_1$. Для $I_2$ получаем то же, но с коэффициентом $c_2$. Из ортогональности $I_1$ и $I_2$ следует $c_1 c_2 = 0$, то есть либо $I_1 = 0$, либо $I_2 = 0$. Получаем противоречие. Следовательно, идемпотент $I_V$ примитивен.

Матричное представление данного идемпотента — квадратная матрица $2^m \times 2^m$ с единицей в верхнем углу. Будем называть это представление каноническим. Форму идемпотента $I_V$ из (2) также будем называть канонической.

Покажем, что в рассматриваемом модуле все эрмитовы примитивные идемпотенты эквивалентны $I_V$, причем изоморфны и соответствующие им группы Клиффорда и спинорные пространства. Пусть имеется эрмитовый примитивный идемпотент $I$. Рассмотрим его матричное представление, с которым его можно отождествить при конкретном выборе матричного представления базисных клиффордовых векторов. Любая эрмитова матрица $A$ приводится к диагональной форме преобразованием эквивалентности с помощью некоторой унитарной матрицы. Приведем матрицу $I$ к диагональной форме таким преобразованием $S_1 I S_1^{-1}$ с помощью некоторой матрицы $S_1$. Поскольку $I^2 = I$, на этой диагонали могут находиться только единицы и нули. Если отличен от нуля больше чем один элемент $I_{ii}$, разложим матрицу $I$ на слагаемые $M_i$, $I = \sum_i M_i$ в которых у $M_i$ ненулевым является только один элемент, стоящий на диагонали в позиции с индексом $i,i$, соответствующей ненулевому элементу $I_{ii}$. Очевидно, что $M_i M_j = M_j M_i = 0$ при $i \neq j$, и $M_i^2 = M_i$. Следовательно, элементы $M_i$ являются идемпотентами. Поэтому идемпотент $I$ раскладывается на сумму ненулевых ортогональных идемпотентов, и он не может быть примитивным. Получили противоречие. Следовательно, после приведения матрицы примитивного эрмитового идемпотента к диагональной форме на ее диагонали имеется только один ненулевой

элемент – единица. Пусть индекс этого элемента $i,i$. Преобразование эквивалентности с помощью матрицы $S_2$, у которой $(S_2)_{ii}=0$, остальные диагональные элементы $(S_2)_{jj}=1$ при $j\neq i$, а недиагональные элементы $(S_2)_{1i}=(S_2)_{i1}=1, (S_2)_{jk}=0, j\neq i, k\neq i$, переводит данный диагональный элемент на место с индексом 1, 1, с сохранением остальных элементов матрицы идемпотента равными нулю. Поэтому матрица любого примитивного эрмитова идемпотента может быть приведена к канонической форме. При этом сам идемпотент проделанными преобразованиями переводится в идемпотент $I'=S_2 S_1 I S_1^{-1} S_2^{-1}$. Такая матрица $\widetilde{I}'$ идемпотента $I'$ может быть представлена в виде произведения $\widetilde{I}'=\widetilde{\overline{\theta}}'_1 \widetilde{\theta}'^1 \widetilde{\overline{\theta}}'_2 \widetilde{\theta}'^2 ... \widetilde{\overline{\theta}}'_m \widetilde{\theta}'^m$ из $m$ матриц $\widetilde{\theta}'^\alpha$ и $\widetilde{\overline{\theta}}'_\alpha$ в канонической форме, соответствующих элементам $\theta'^\alpha$ и $\overline{\theta}'_\alpha$. Следовательно, соответствующее равенство справедливо и для элементов алгебры Клиффорда, представлениями которых служат данные матрицы:

$$I'=\overline{\theta}'_1 \theta'^1 \overline{\theta}'_2 \theta'^2 ... \overline{\theta}'_m \theta'^m. \qquad (3)$$

Введем величины $e_{2\alpha-1}'=s_{2\alpha-1}(\theta'^\alpha+\overline{\theta}'_\alpha)$, $e_{2\alpha}'=is_{2\alpha}(\theta'^\alpha-\overline{\theta}'_\alpha)$. Поскольку для величин $e_k'$, $e_l'$ выполняются те же соотношения $\{e_k',e_l'\}=(s_k)^2 \delta_l^k$, что и соотношения $\{e_k,e_l\}=(s_k)^2 \delta_l^k$ для $e_k$, $e_l$, по обобщенной теореме Паули [3] найдется единственный унимодулярный элемент $S$, обеспечивающий преобразование эквивалентности $e_k'=S e_k S^{-1}$. Но это преобразование алгебры Клиффорда с базисными векторами $e_k$ в изоморфную ей алгебру Клиффорда с базисными векторами $e_k'$. Поэтому идемпотент (3) – это идемпотент в канонической форме (2) с канонической формой его матричного представления. Таким образом, для произвольного эрмитового примитивного идемпотента $I$ найдется невырожденный унимодулярный элемент алгебры S, с помощью которого осуществляется преобразование эквивалентности, переводящее его в идемпотент $I_V$ в канонической форме (2). То есть в рассматриваемом модуле все эрмитовы примитивные идемпотенты эквивалентны $I_V$. Более того, поскольку преобразование эквивалентности переводит базисные клиффордовы векторы первоначальной алгебры в базисные клиффордовы векторы изоморфной ей клиффордовой алгебры, в этих алгебрах изоморфны как группы Клиффорда, так и спинорные пространства, соответствующие этим группам и идемпотентам. Что и доказывает утверждение.

При действии на $I_V$ из (2) величины $\theta^\alpha$ служат операторами рождения, $\overline{\theta}_\alpha$ – операторами уничтожения, а $I_V$ служит для них клиффордовым вакуумом.

В работе [1] показано, что комплексная алгебра $Cl(2m)$ разложима на прямую сумму $2^m$ минимальных взаимно ортогональных левых идеалов, порождаемых базисом Витта. В комплексном модуле над вещественной алгеброй Клиффорда допустимо аналогичное разложение с использованием фермионных переменных. Пусть $K_\alpha=\overline{\theta}_\alpha \theta^\alpha$ и $\widetilde{K_\alpha}=\theta^\alpha \overline{\theta}_\alpha$. Поскольку $K_\alpha+\widetilde{K_\alpha}=1$, имеем $\prod_{\alpha=1..m}(K_\alpha+\widetilde{K_\alpha})=1$, и, раскладывая произведение на сумму слагаемых, получаем $\sum_{i=1..2^m} I_i =1$, где $I_1=K_1 K_2 ... K_m$, $I_2=\widetilde{K_1} K_2 ... K_m$, $I_3=K_1 \widetilde{K_2} ... K_m$, и т.д. В каждом из слагаемых $I_i$ m множителей, в качестве которых выступают либо $K_\alpha$, либо $K_\alpha\tilde{\ }$, т.е. 2 варианта в каждой из m позиций. Поскольку $K_\alpha K_\alpha\tilde{\ } = K_\alpha\tilde{\ } K_\alpha = 0$, $I_i I_j = I_j I_i = 0$ при $i \neq j$ и $(I_i)^2 = I_i$. То есть это набор ортогональных идемпотентов. Поскольку $I_1 = I_V$, данный идемпотент примитивный. Все $I_i$ изоморфны и отличаются только заменой $\overline{\theta}_\alpha$ и $\theta^\alpha$ в соответствующей паре, поэтому они также примитивны. Следовательно, идеалы, образованные с их помощью и образующие разложение модуля на ортогональные пространства спиноров, минимальны. Что завершает доказательство.

Каждый из $2^m$ идемпотентов $I_i$ является клиффордовым вакуумом, приводимым к каноническому виду (2) преобразованием подобия. Однако одновременно к каноническому виду может быть приведен только один из $2^m$ клиффордовых вакуумов. Это приводит к необходимости одновременно рассматривать не один, а сразу несколько вакуумов. В половине из клиффордовых вакуумов модуля на первой позиции стоит $\overline{\theta}_1\theta^1$, и для них $\theta^1$ будет служить оператором рождения, а $\overline{\theta}_1$ – оператором уничтожения, а в остальных – $\overline{\theta}_1$ оператор рождения, а $\theta^1$ – оператор уничтожения. Таким образом, привычные представления об однозначности роли оператора рождения или уничтожения, применимые в одном клиффордовом вакууме, необходимо заменить на более сложные.

В матричном представлении модулю соответствует квадратная матрица $2^m \times 2^m$, каждому из спинорных пространств, порожденных клиффордовым вакуумом – столбец этой матрицы, а самому клиффордову вакууму в этом столбце соответствует единичный элемент, стоящий на диагонали матрицы модуля. Произвольный элемент $\Psi$ идеала, образованного с помощью $I_1$, может быть представлен в виде

$$\Psi = (\psi_0 + \psi_{k_1}\theta^{k_1} + \psi_{k_1 k_2}\theta^{k_1}\theta^{k_2} + \psi_{k_1 k_2 k_3}\theta^{k_1}\theta^{k_2}\theta^{k_3} + ... + \psi_{123...m}\theta^1\theta^2...\theta^m)I_V, \qquad (4)$$

где по повторяющимся индексам идет суммирование, а $\psi_0$, $\psi_{k_1}$, $\psi_{k_1 k_2}$, ..., $\psi_{k_1 k_2...k_m}$ – числовые коэффициенты, преобразующиеся как компоненты спинора. Элементы разложения (4) нельзя считать 0-частичными, 1-частичными, и т.д. Они все являются компонентами одночастичного состояния – спинора. Для остальных идеалов разложения аналогичны с взаимной заменой $\theta^k$ и $\overline{\theta}_k$ для соответствующего $k$. $2^m$ левых идеалов модуля, порожденных идемпотентами $I_i$, можно сопоставить $2^m$ фермионам.

Таким образом, с помощью фермионных переменных можно построить клиффордовы вакуумы, векторы состояния фермионов и образующие клиффордовой алгебры. Поэтому фермионные переменные могут рассматриваться как более фундаментальные математические объекты, чем спиноры.

### Список литературы